\documentclass{PoS}

\title{Isolated Leptons and Single Top at HERA}

\ShortTitle{Isolated Leptons and Single Top at HERA}

\author{\speaker{David M. South}\thanks{on behalf of the H1 and ZEUS Collaborations}\\
        Technische Universit\"{a}t Dortmund\\
        Experimentelle Physik V\\
        44221 Dortmund, Germany\\
        E-mail: \email{david.south@desy.de}}

\abstract{The search for events containing isolated leptons
and missing transverse momentum produced in $e^{\pm}p$
collisions is performed individually and in a common phase space
with the H1 and ZEUS detectors at HERA.
The combined H1 and ZEUS data sample corresponds to an integrated
luminosity of $0.98$~fb$^{-1}$ and comprises the complete high
energy data from the HERA programme.
A total of $81$ events are observed in the data, compared to a
Standard Model prediction of $87.8 \pm 11.0$, which is dominated
by single $W$ production.
At large hadronic transverse momentum $P_{T}^{X} >$~25~GeV in
the $e^{+}p$ data, integrated luminosity $0.58$~fb$^{-1}$, $23$ data events are
observed compared to a SM prediction of $14.0 \pm 1.9$.
The total single $W$ production cross section is measured as
$1.06 \pm 0.16 ({\rm stat.}) \pm 0.07 ({\rm sys.})$ in agreement
with a SM expectation of $1.26 \pm 0.19$.
The isolated lepton events are examined in the context of a search for
anomalous single top production, where the hadronic decays of the $W$
are additionally considered.
Although several top--like candidates are present in the H1 data no clear
signal is observed and an upper limit on the anomalous single top production
cross section of $\sigma_{ep\rightarrow etX} < 0.25$~pb is established at the
$95\%$ confidence level. This limit corresponds to an upper bound on
the anomalous magnetic coupling of $\kappa_{tu\gamma} < 0.18$
assuming a top mass of $175$~GeV.
}

\FullConference{XVIII International Workshop on Deep-Inelastic Scattering and Related Subjects, DIS 2010\\
		April 19-23, 2010\\
		Firenze, Italy}

\begin{document}

\section{Introduction}

At HERA, protons with an energy up to $920$~GeV were brought into
collision with electrons or positrons of energy $27.6$~GeV at two
experiments, H1 and ZEUS, each of which collected about
$0.5$~fb$^{-1}$ of data in the period $1994$--$2007$.
The collisions produced at HERA at a centre of mass energy of
up to $319$~GeV provide an ideal environment to study rare processes,
set constraints on the Standard Model (SM) and search for new
particles and physics beyond the Standard Model (BSM).
Events containing isolated leptons and missing transverse momentum
in the final state may be a signature of rare processes and such events
have been observed at HERA~\cite{h1isolep1998,zeusisolep2000,h1isolep2003}.
An excess of data events compared to the SM prediction at large hadronic
transverse momentum $P_{T}^{X}$ was observed by the H1
Collaboration~\cite{h1isolep2003} in an analysis of
the HERA~I data ($1994$--$2000$), which was mostly $e^{+}p$ collisions.
This was not confirmed by the ZEUS Collaboration in a more restricted
phase space~\cite{zeustop2003}.
The analyses of the complete H1~\cite{h1isolep2009} and
ZEUS~\cite{zeusisolep2009} data have now been finalised, as well
as a combined analysis perfomed in a common phase
space~\cite{h1zeusisolep2010}, which makes use of the full
$0.98$~fb$^{-1}$ of HERA data.
Furthermore, the selected events have been examined in the context of
anomalous single top production via a flavour changing neutral
current (FCNC) interaction.


\section{Events with Isolated Leptons and Missing Transverse Momentum }
\label{sec:isolep}

The main SM contribution to the signal\footnote{Processes are defined
  as {\it signal} if they produce events containing a high $P_{T}$ isolated,
  electron or muon and at least one high $P_{T}$ neutrino, which escapes
  detection and leads to $P_{T}^{\rm miss}$ in the final state.}
topology is the production of real $W$ bosons via photoproduction with
subsequent leptonic decay
$ep\rightarrow eW^{\pm}$($\rightarrow \ell\nu$)$X$, where the hadronic
system $X$ has typically low transverse momentum $P_{T}$.
The equivalent charged current (CC) process
$ep \rightarrow \nu$$W^{\pm}$($\rightarrow \ell\nu$)$X$, also contributes to the
total signal rate, although only at a level of about 7\%.
The production of $Z^{0}$ bosons with subsequent decay to neutrinos
$ep \rightarrow eZ^{0}$($\rightarrow \nu\bar{\nu}$)$X$ results in a
further minor contribution\footnote{This process is not included in
the ZEUS analysis.} to the total signal rate in the electron
channel at a level of 3\%.
SM background enters the electron channel due to mismeasured neutral
current (NC) deep inelastic scattering (DIS) events and the muon
channel due to lepton pair (LP) events in which one muon escapes
detection, both cases resulting in apparent (fake) missing
transverse momentum.
CC DIS background, which contains intrinsic missing transverse
momentum, enters the final sample in both lepton channels, where a final
state particle is interpreted as the isolated electron or muon.


The common event selection is based on those used by H1~\cite{h1isolep2009}
and ZEUS~\cite{zeusisolep2009}.
Lepton candidates are required to have $P_{T}^{\ell}>10$~GeV,
to be in the central region of the detector $15^{\circ} < \theta_{\ell} <
120^{\circ}$ and to be isolated from tracks and identified jets in the event.
The event must also exhibit significant missing transverse momentum,
$P_{T}^{\rm miss}>12$~GeV.
Further topological and kinematic cuts are applied to reject the 
remaining SM background.
A full description of the event selection is presented in~\cite{h1zeusisolep2010}.

\begin{figure*}[t]
\centering
\includegraphics[width=0.5\columnwidth]{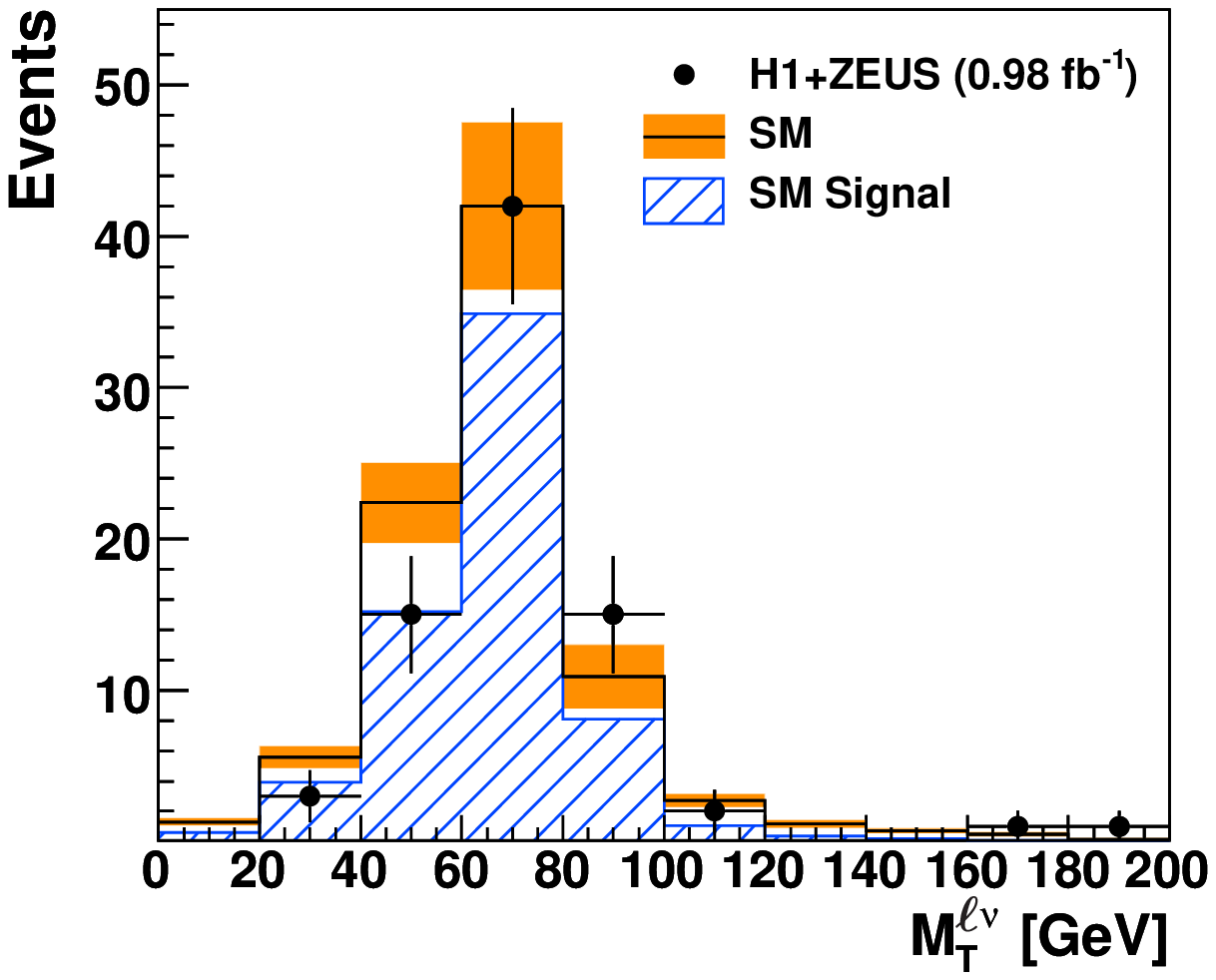}
\includegraphics[width=0.47\columnwidth]{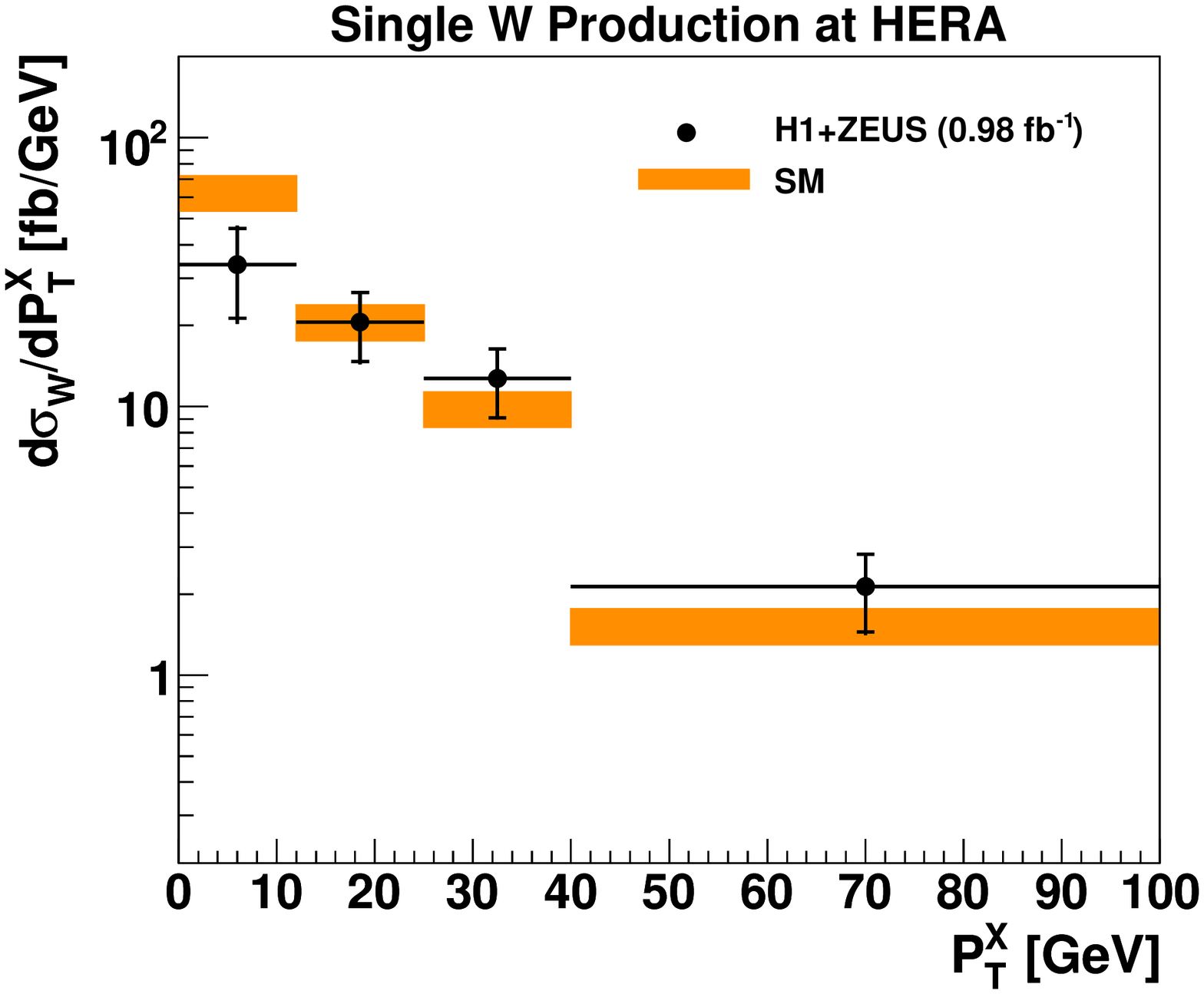}
\vspace{-0.35cm}
\caption{Left: The lepton--neutrino transverse mass $M_{T}^{\ell\nu}$
  of events with an isolated electron or muon and missing transverse
  momentum. The data (points) are compared to the SM
  expectation (open histogram). The signal component of the SM
  expectation, dominated by single $W$ production, is shown as the
  striped histogram. The total uncertainty on the SM expectation is
  shown as the shaded band. Right: The single $W$ production
  cross section as a function of the hadronic transverse momentum,
  $P_{T}^{X}$. The inner error bar represents the statistical error
  and the outer error bar indicates the statistical and
  systematic uncertainties added in quadrature. The shaded band
  represents the uncertainty on the SM prediction.}
\label{fig:isolep}
\end{figure*}

The results of the analysis of the complete $e^{\pm}p$ HERA data are
shown in table~\ref{tab:isolep}.
In general, a good agreement is observed between the data and the SM
prediction, where the signal component forms the main part of the expectation.
The lepton--neutrino transverse mass distribution is shown in
figure~\ref{fig:isolep}~(left) and displays the Jacobian peak
expected from single $W$ production.
For $P_{T}^{X}>25$~GeV, $29$ events are observed in the data, compared
to a SM prediction of $24.0 \pm 3.2$ for the complete HERA $e^{\pm}p$
data.
In the $e^{+}p$ data alone, where a small excess of data is seen in
the H1 analysis~\cite{h1isolep2009}, $23$ events are observed in the data,
compared to a SM prediction of $14.0 \pm 1.9$.

\begin{table}[t]
\begin{center}
\begin{tabular}{ c c c c c c }
Channel & & Data & SM & Signal (W) & Background\\
\hline
Electron & Total   & $61$ & $69.2 ~\pm~ ~~8.2$ & $48.3 ~\pm~ 7.4$ & $20.9 ~\pm~ 3.2$ \\
& $P^X_T > 25$ GeV & $16$ & $13.0 ~\pm~ ~~1.7$ & $10.0 ~\pm~ 1.6$ &  ~~$3.1 ~\pm~ 0.7$ \\
\hline
Muon     & Total   & $20$ & $18.6 ~\pm~ ~~2.7$ & $16.4 ~\pm~ 2.6$ &  ~~$2.2 ~\pm~ 0.5$ \\
& $P^X_T > 25$ GeV & $13$ & $11.0 ~\pm~ ~~1.6$ &  ~~$9.8 ~\pm~ 1.6$ &  ~~$1.2 ~\pm~ 0.3$ \\
\hline
Combined & Total   & $81$ & $87.8 ~\pm~ 11.0$  & $64.7 ~\pm~ 9.9$ & $23.1 ~\pm~ 3.3$ \\
& $P^X_T > 25$ GeV & $29$ & $24.0 ~\pm~ ~~3.2$  & $19.7 ~\pm~ 3.1$ &  ~~$4.3 ~\pm~ 0.8$ \\
\hline                                        
\end{tabular}
\end{center}
\vspace{-0.35cm}
\caption{Observed and predicted number of events with an
  isolated electron or muon and missing transverse momentum
  in $0.98$~fb$^{-1}$ of HERA data. The results are shown for the full selected
  sample and for a subsample with $P_{T}^{X}>25$~GeV.
  The signal, total SM and background contributions are shown.
  The quoted errors contain statistical and systematic
  uncertainties added in quadrature.}
\label{tab:isolep}
\end{table}

The total and differential single $W$ production cross sections are
evaluated from the number of observed events, subtracting the number
of background events and taking into account the acceptance and 
integrated luminosity of the two experiments.
The single $W$ cross section is measured as
$1.06 \pm 0.16 ({\rm stat.}) \pm 0.07 (\rm sys.)$~pb, in agreement
with the SM prediction of $1.26 \pm 0.19$~pb.
The cross section is also measured as a function of $P_{T}^{X}$,
as shown in figure~\ref{fig:isolep}~(right).


Further studies of events with isolated leptons and missing
transverse momentum are performed by H1~\cite{h1isolep2009}.
A complementary search for events with an isolated tau
lepton and missing $P_{T}$ is done by selecting tau candidates,
identified as narrow hadronic jets resulting from $1$--prong decays,
in coincidence with large missing transverse momentum.
In the final event sample, $18$ events are selected, compared to a SM
expectation of $23.2 \pm 3.8$, which is dominated by CC DIS
background.
Additionally, the $W$ cross section measured by H1 is used to derive
single parameter limits on the $WW\gamma$ gauge coupling
parameters $\Delta\kappa$ and $\lambda$ at $95$\%~CL.
Finally, the $W$ polarisation fractions are measured for the
first time at HERA by H1 and found to be consistent with the
SM expectation.

\section{Search for Anomalous Single Top Production}
\label{sec:top}

The production of single top quarks is kinematically possible via the
$ep$ collisions at HERA due to the large centre of mass energy, although the
dominant SM process has a negligible cross section of less than~$1$~fb.
However, in several extensions of the SM the top quark is predicted to undergo
Flavour Changing Neutral Current (FCNC) interactions, which could lead to a sizeable
anomalous single top production cross section at HERA~\cite{toptheory}.
The interaction of a top quark with $U$-type quarks via a photon is
described in the effective Lagrangian by a magnetic coupling $\kappa_{tU\gamma}$.
Dedicated searches for anomalous single top production have
been performed at HERA~I, where no conclusive signal
was observed~\cite{h1top2003,zeustop2003} and the
HERA~II data have now been analysed~\cite{h1top2009,zeustop2009}.


The top quark is detected via its decay $t \rightarrow b W^{+}$.
In the case of leptonic $W$ decays, the signature is the same as the
topology described in section~\ref{sec:isolep}, where the hadronic
final state exhibits high $P_{T}^{X}$ originating from the fragmentation
of the $b$ quark, thus providing a possible explanation of the
data excess observed by the H1 experiment\footnote{It should
be noted that single top production cannot explain the observed difference
between the $e^{+}p$ and $e^{-}p$ data.}.
In the case of the hadronic $W$ boson decays, the signature of
single top quark production consists of three high $P_{T}$ jets
with an invariant mass compatible with the top quark mass.

\begin{figure*}[h]
\centering
\includegraphics[width=0.475\columnwidth]{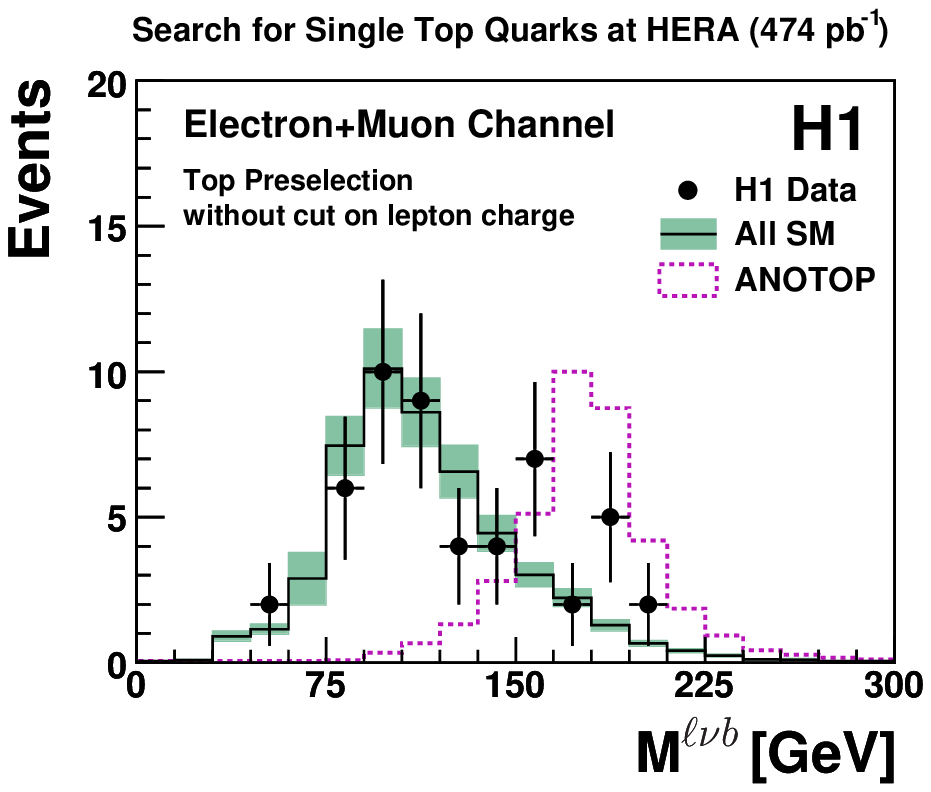}
\includegraphics[width=0.475\columnwidth]{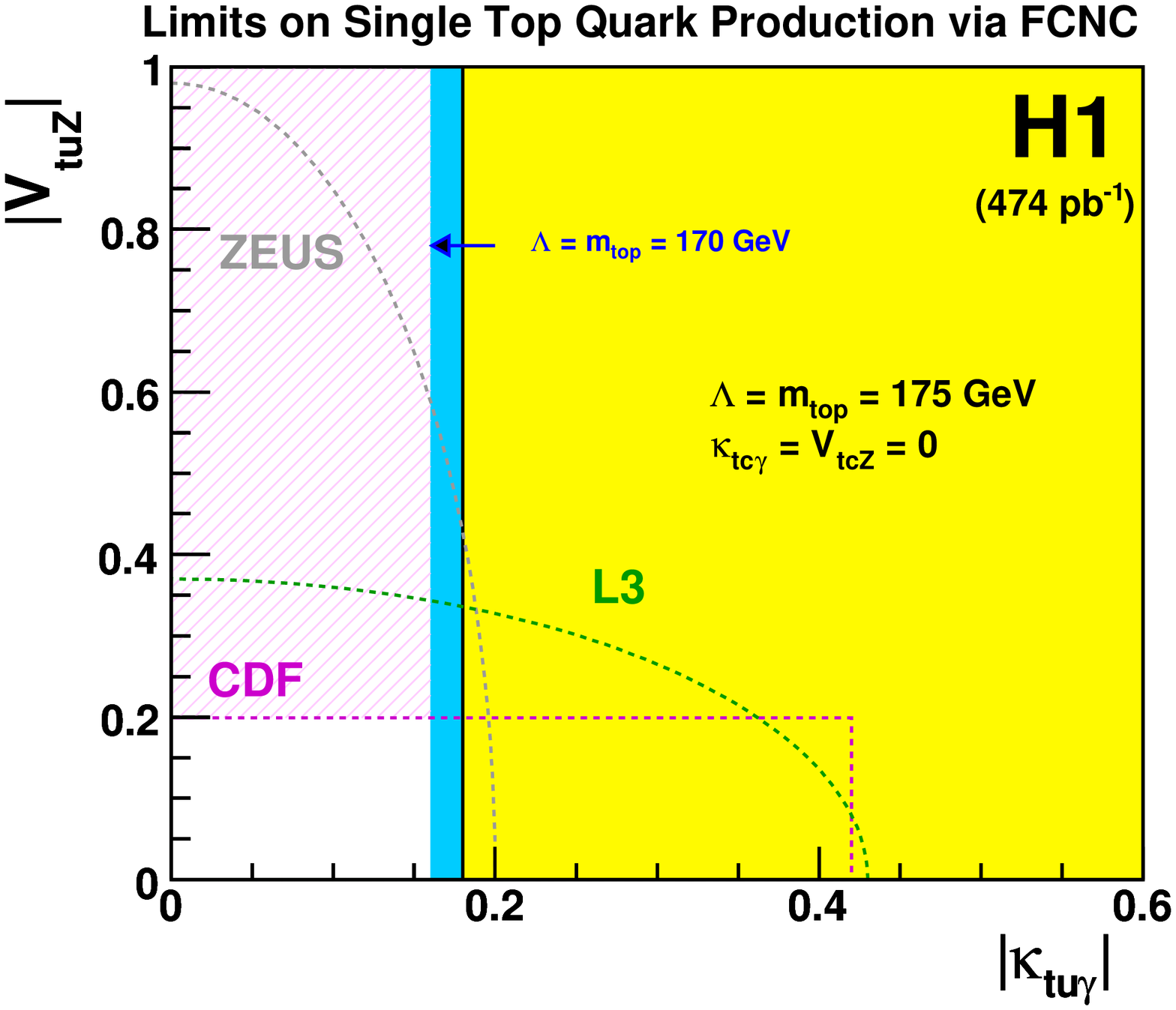}
\vspace{-0.35cm}
\caption{Left: Distribution of the reconstructed top mass
  $M^{\ell\nu b}$ in the electron and muon channels after neutrino reconstruction
  but before the cut on the lepton charge. The data are the points,
  the total SM expectation as the open histogram. The  total
  uncertainty on the SM is shown as the shaded band. The
  single top prediction is shown with arbitrary normalisation
  (dashed histogram). Right: Various exclusion limits at $95$\% CL on the anomalous
  $\kappa_{tu\gamma}$ and $V_{tuZ}$ couplings for a top mass of
  $175$~GeV, obtained at HERA (H1~\cite{h1top2009} and
  ZEUS~\cite{zeustop2003} experiments), at the Tevatron
  (CDF experiment~\cite{cdftop,cdftop2}) and at LEP
  (L3 experiment~\cite{l3top}).
  Anomalous couplings of the charm quark are neglected
  $\kappa_{tc\gamma}=V_{tcZ}=0$, as are vector couplings to
  the $Z^0$ boson $v_{tUZ}=0$ in the case of the H1 limit.}
\label{fig:top}
\end{figure*}

An H1 publication~\cite{h1top2009} examines both the leptonic
and hadronic decay channels, where for the former the selection
is based on the full sample of events selected in the isolated
leptons analysis~\cite{h1isolep2009}.
The distribution of the reconstructed top mass $M^{\ell\nu b}$ in the leptonic channels
after neutrino reconstruction, but before a requirement is made on a positive lepton
charge, is shown in figure~\ref{fig:top} (left).
In order to separate a potential FCNC top signal from the SM background, both the
leptonic and hadronic decay channels combine relevant event level observables
into a multi-variate discriminant, trained using single top Monte Carlo (MC) as
the model and $W$ MC as the SM background. 
In the absence of a top signal, limits on the signal cross section
are extracted from the discriminator spectra using a maximum likelihood
method~\cite{h1top2009}.
An upper bound on the cross section of $\sigma_{ep \rightarrow e t X}<$~0.25~pb at 95\% CL
is found, which is translated into an upper bound on the coupling
$\kappa_{tu\gamma}<$~0.18 for a top mass $m_{top}=175$~GeV.
This limit is shown in figure~\ref{fig:top} (right) in comparison to
results from other colliders~\cite{cdftop,cdftop2,l3top}.
A preliminary result from ZEUS using HERA~II data~\cite{zeustop2009}, which
examines only the leptonic $W$ decay channel and is not shown in
figure~\ref{fig:top} (right), sets limits of
$\sigma_{ep\rightarrow etX} < 0.13$~pb and
$\kappa_{tu\gamma} < 0.13$ for a top mass $m_{\rm top}=171.2$~GeV.

\section{Conclusions}

Analyses of events with isolated leptons with $P_{T}^{\rm miss}$ have
recently been published individually and as a combined analysis by H1 and ZEUS.
In general, a good agreement is observed with the SM and
the single $W$ production cross section is measured as
$\sigma_{W} = 1.06 \pm 0.16 ({\rm stat.}) \pm 0.07 (\rm sys.)$~pb.
Searches for FCNC single top production are also performed and in the
absence of a signal an upper limit on the cross section and anomalous
coupling $\kappa_{tU\gamma}$ are derived.

\end{document}